\documentclass[amsmath,amssymb,aps,prb,twocolumn,superscriptaddress]{revtex4-2}
\usepackage{graphicx}
\usepackage{float}
\usepackage{color}

\begin{document}
\title{Robust Coulomb Gap and Varied-temperature Study of Epitaxial 1T'-WSe$_2$ Monolayers}

\author{Wang Chen}
\thanks{These two authors contributed equally to this work}
\affiliation{National Laboratory of Solid State Microstructure, School of Physics, Nanjing University, Nanjing, 210093, China}
\author{Mengli Hu}
\thanks{These two authors contributed equally to this work}
\affiliation{Department of Physics, Hong Kong University of Science and Technology, Clear Water Bay, Hong Kong, China}
\author{Junyu Zong}
\affiliation{National Laboratory of Solid State Microstructure, School of Physics, Nanjing University, Nanjing, 210093, China}
\author{Xuedong Xie}
\affiliation{National Laboratory of Solid State Microstructure, School of Physics, Nanjing University, Nanjing, 210093, China}
\author{Wei Ren}
\affiliation{Vacuum Interconnected Nanotech Workstation (Nano-X), Suzhou Institute of Nano-Tech and Nano-Bionics (SINANO), Chinese Academy of Sciences, Suzhou, 215123, China}
\author{Qinghao Meng}
\author{Yu Fan} 
\author{Qichao Tian} 
\author{Shaoen Jin} 
\author{Xiaodong Qiu} 
\author{Kaili Wang}
\affiliation{National Laboratory of Solid State Microstructure, School of Physics, Nanjing University, Nanjing, 210093, China}
\author{Can Wang}
\affiliation{National Laboratory of Solid State Microstructure, School of Physics, Nanjing University, Nanjing, 210093, China}
\affiliation{School of Physics and Electronic Sciences, Changsha University of Science and Technology, Changsha 410114, China}
\author{Junwei Liu}
\thanks{To whom correspondence should be addressed: liuj@ust.hk, lwang2017@sinano.ac.cn, zhangyi@nju.edu.cn}
\affiliation{Department of Physics, Hong Kong University of Science and Technology, Clear Water Bay, Hong Kong, China}
\author{Fang-Sen Li}
\author{Li Wang}
\thanks{To whom correspondence should be addressed: liuj@ust.hk, lwang2017@sinano.ac.cn, zhangyi@nju.edu.cn}
\affiliation{Vacuum Interconnected Nanotech Workstation (Nano-X), Suzhou Institute of Nano-Tech and Nano-Bionics (SINANO), Chinese Academy of Sciences, Suzhou, 215123, China}
\author{Yi Zhang}
\thanks{To whom correspondence should be addressed: liuj@ust.hk, lwang2017@sinano.ac.cn, zhangyi@nju.edu.cn}
\affiliation{National Laboratory of Solid State Microstructure, School of Physics, Nanjing University, Nanjing, 210093, China}
\affiliation{Collaborative Innovation Center of Advanced Microstructures, Nanjing University, Nanjing, 210093, China}
\affiliation{Hefei National Laboratory, Hefei 230088, China}
 
\begin{abstract}
The transition metal dichalcogenides (TMDCs) with a 1T' structural phase are predicted to be two-dimensional topological insulators at zero temperature. 
Although the quantized edge conductance of 1T'-WTe$_2$ has been confirmed to survive up to 100 K \cite{QSHI_TMDCs3}, this temperature is still relatively low for industrial applications. 
Addressing the limited studies on temperature effects in 1T'-TMDCs, our research focuses on the electronic and crystal properties of the epitaxial 1T'-WSe$_2$ monolayers grown on bilayer graphene (BLG) and SrTiO$_3$(100) substrates at various temperatures.
For the 1T'-WSe$_2$ grown on BLG, we observed a significant thermal expansion effect on its band structures with a thermal expansion coefficient of $\sim$60$\times$10$^{-6}$ K$^{-1}$.
In contrast, the 1T'-WSe$_2$ grown on SrTiO$_3$(100) exhibits minimal changes with varied temperatures due to the enhanced strain exerted by the substrate. 
Besides, A significant Coulomb gap (CG) was observed pinned at the Fermi level in the angle-resolved photoemission spectroscopy (ARPES) and scanning tunneling spectroscopy (STS). 
The CG was founded to decrease with increasing temperatures, and can persist up to 200 K for 1T'-WSe$_2$/BLG, consistent with our Monte Carlo simulations.
The robustness of the CG and the positive fundamental gap endow the epitaxial 1T'-WSe$_2$ monolayers with huge potential for realizing the quantum spin Hall devices. 
\end{abstract}

\maketitle
\section{Introduction}

The quantum spin Hall (QSH) state, also known as a two-dimensional (2D) topological insulator (TI), is a remarkable quantum phenomenon characterized by a pair of helical edge state counterpropagating without back-scattering \cite{QSHI_gra,QSHI_zsc,QSHI_TI}. 
The transition metal dichalcogenides (TMDCs) monolayers with a 1T' structural phase are a typical family of 2D TIs \cite{QSHI_TMDCs1, QSHI_TMDCs2}. 
Among all the TMDCs, 1T'-WTe$_2$ monolayer has been confirmed to exhibit quantized edge conductance up to 100 K \cite{QSHI_TMDCs3}.
Although the working temperature is the record for QSH effect, it still requires largely improvement for practical applications. 
Therefore, the influence of temperature on the electronic properties of TMDCs monolayers is a crucial area for further investigations.
Another challenge is the short quantized edge channel in 1T'-WTe$_2$ \cite{QSHI_TMDCs3}.
One possible solution is exploring other TMDCs with a single-particle positive band gap. 
Among the six predicted TI compounds of TMDCs, 1T'-$MX_2$ ($M$ = Mo, W; $X$ = S, Se, Te) \cite{QSHI_TMDCs1,QSHI_WTe2_1,WTe2_ARPES,QSHI_TMDCs2,QSHI_WTe2_2,WSe2_STM,WSe2_ARPES}, four, including MoS$_2$, WS$_2$, MoSe$_2$, and WSe$_2$, possess a positive band-gap but are metastable thermodynamically in the 1T' phase \cite{QSHI_TMDCs1,WSe2_transition5}. 
Recent developments in 1T' sample fabrication have enabled the preparation of 1T'-WSe$_2$ on graphene or SrTiO$_3$(100) substrate using molecular beam epitaxial (MBE) \cite{WSe2_ARPES,WSe2_transition5,WSe2_interface}. 
The relatively large band gap of 1T'-WSe$_2$ \cite{WSe2_STM,WSe2_interface} makes it a promising candidate for practical realization in QSH devices.
\par

2D materials possess unique properties and edges in constructing feasible and novel heterostructures for their weak van der Waals interface/surface interactions \cite{2D_review,2D_hetero,Gra_2D,Review_2D_graphene}. 
For example, heterojunctions consisting of TIs and superconductors or magnetic thin films can facilitate Majorana zero modes and the quantum anomalous Hall effect, respectively \cite{TI_MZM,MZM_TS,SC_Proximity,QAHE_prediction,QAH_MTI}. 
Moreover, due to their low carrier density and reduced dimensionality, 2D materials exhibit enhanced Coulomb interaction and correlation effects. 
This is exemplified by the giant exciton binding energy reported in 1H-MoS$_2$ and 1H-WSe$_2$ \cite{MoSe2_exciton,WSe2_thinfilm}, and the discovery that 1T'-WTe$_2$ is an exciton insulator in the clean limit with a large binding energy \cite{Exciton_1,Exciton_2}. 
Furthermore, it has been found that 1T'-WTe$_2$ and Pb film host the Coulomb gap (CG) with a diminishing density of states at Fermi level \cite{WTe2_coulomb,Coulomb_STO}, and superconductivity can emerge under gate voltages in 1T'-WTe$_2$ \cite{WTe2_SC1,WTe2_SC2}. 
Differing from the spin-orbital-coupling (SOC) induced band insulator of TMDCs, the CG is introduced by the interaction of quasi-particles at Fermi level \cite{WTe2_coulomb,Exciton_2}. 
These characteristics make the 1T'-TMDCs a promising platform for studying interaction and correlation effects in addition to the topological physics in 2D system.
\par

In this article, we investigate the electronic structures of epitaxial 1T'-WSe$_2$ monolayers grown on bilayer graphene (BLG) and SrTiO$_3$(100) substrates (denoted as 1T'-WSe$_2$/BLG, and 1T'-WSe$_2$/SrTiO$_3$) under varied temperatures using in-situ angle-resolved photoemission spectroscopy (ARPES) and scanning tunneling microscopy/spectroscopy (STM/STS). 
With increasing temperatures, we observed significant thermal expansion in 1T'-WSe$_2$/BLG via STM. 
Our combined ARPES detection and first-principle calculations demonstrate the non-significant alteration in band structure. 
Besides, we observed robust CGs pinned at Fermi level in both cases, with gap sizes decreasing as temperatures increased. 
Notably, the CG of 1T'-WSe$_2$/BLG can persist at high temperature of 200 K.
The CG of 1T'-WSe$_2$/SrTiO$_3$ is relatively smaller due to the distinct heavier doping levels that provide stronger screening on the Coulomb interaction.
Our Monte Carlo numerical simulations, incorporating variables of temperature and doping effects on CG, provide further insight and successfully reproduce the observed CG evolution in our experiments.
Overall, the persistence of CG, which suppresses the density of state (DOS), enhances the insulating properties of 1T'-WSe$_2$. 
Our findings suggest that 1T'-WSe$_2$ is a promising candidate for QSH device and unveil the detailed temperature effect on its electronic structure, opening avenues for further exploration of interaction and correlation research in 2D system.
\par

\onecolumngrid
\begin{figure}[h]
\centering
\begin{minipage}{\textwidth}
\includegraphics{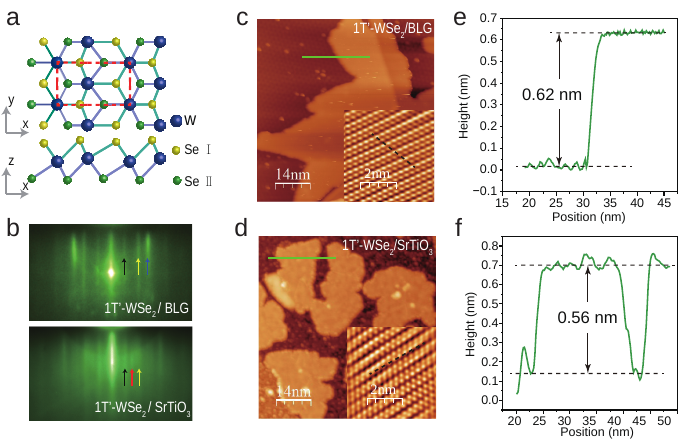}
\caption{
\label{figure1}
\textbf{(a)} Top and side perspective illustrations of monolayer 1T'-WSe$_2$ lattice structure.
Se I and Se II represent the Se atoms at the top surface and bottom interface, respectively.
\textbf{(b)} RHEED patterns obtained from the 1T'-WSe$_2$ monolayers grown on BLG (upper image) and SrTiO$_3$(001) (lower image). 
The black and yellow arrows indicate the diffraction stripes from $\frac{1}{2}\bm{a}$ and $\bm{b}$ of 1T'-WSe$_2$ lattice. 
The blue and red arrows indicate the  diffraction stripes from the BLG and SrTiO$_3$ substrates, respectively. 
\textbf{(c)} and \textbf{(d)} STM topographic images of WSe$_2$ monolayer islands on (c) BLG substrate (denoted as 1T'-WSe$_2$/BLG) and (d) SrTiO$_3$(001) substrate (denoted as 1T'-WSe$_2$/SrTiO$_3$), respectively.
Scanning parameters include a bias voltage $U$ = +1.0 V, a tunneling current $I_t$ = 100 pA, and a temperature $T$ approximately 4 K.
Insets present atomic resolution images of the 1T'-WSe$_2$ monolayers.
\textbf{(e)} and \textbf{(f)} Height profiles along the green lines shown in (c) and (d), respectively.
}
\end{minipage}
\end{figure}
\twocolumngrid
\par

\section{Results}
\subsection{Characterization of the Epitaxial 1T'-WSe$_2$}
Our varied-temperature study focuses on epitaxial 1T'-WSe$_2$ grown on two substrates: BLG and SrTiO$_3$(100). The successful growth of 1T'-WSe$_2$ on both substrates is evidenced in RHEED and STM topography. 
Figure~\ref{figure1}(a) shows the atomic structure of 1T'-WSe$_2$: Se I - W - Se II atomic layers form an A-B-C stacking structure, and W atoms distort in the $x-z$ plane, distinguishing the 1T' phase from 1H and 1T phases. 
Compared to the 1H phase, the distinctive streak features of the 1T'-WSe$_2$ structure, as shown in the atomic resolution images from STM [Fig.~\ref{figure1}(c, d)], can be attributed to the approximate twice relation between lattice constants and the height variations between the nearest Se I atoms along the $y$ direction. 
This lattice constants relation in 1T'-WSe$_2$ is also evident in the diffraction patterns from RHEED as shown in Fig.~\ref{figure1}(b): the distance from the central point indicated by the black arrow is approximately half that of the yellow arrow. 
The diffraction patterns denoted by blue (in the upper panel) and red (in the lower panel) arrows are from the BLG and SrTiO$_3$(100) lattices, respectively.
We also present the height profile of the grown 1T'-WSe$_2$ films in Fig.~\ref{figure1}(e,d). 
The 1T'-WSe$_2$ grown on BLG substrate shows a height of $\sim$0.62 nm, and the 1T'-WSe$_2$ grown on SrTiO$_3$(100) substrate shows a height of $\sim$0.56 nm. These height values indicate that the thickness of the grown 1T'-WSe$_2$ films are indeed monolayer \cite{WSe2_STM,WSe2_interface,WSe2_transition5}.

\subsection{Thermal Expansion Effects}

\begin{figure}[h]
\begin{minipage}[c]{\textwidth}
    \centering
    \includegraphics{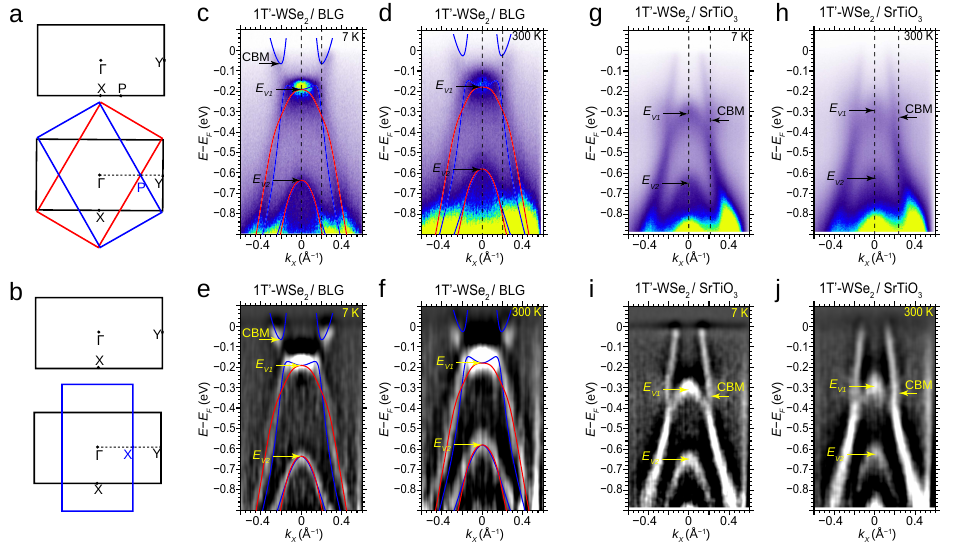}
    \caption{\label{figure2}
    \textbf{(a) and (b)} Brillouin zones (upper) with labeled high symmetry points and complex Brillouin zones (lower) with three domains connected by three-fold rotation and two domains connected by two-fold rotation for 1T'-WSe$_2$/BLG and 1T'-WSe$_2$/SrTiO$_3$, respectively.
    \textbf{(c) and (d)} ARPES spectra of along the Y-$\Gamma$-Y direction taken from 1T'-WSe$_2$/BLG at (c) 7 K and (d) 300 K, respectively. 
    \textbf{(e) and (f)} Second-derivative spectra of the ARPES date in (c) and (d).
    The red and blue solid lines are the calculated band structures of 1T'-WSe$_2$/BLG along the Y-$\Gamma$-Y and P-$\Gamma$-P directions, respectively. 
    \textbf{(g) and (h)} ARPES spectra along the Y-$\Gamma$-Y direction taken from 1T'-WSe$_2$/SrTiO$_3$ at 7 K and 300 K, respectively. 
    \textbf{(i) and (j)} Second-derivatives of the ARPES spectra (g) and (h).    
    }
\end{minipage}
\end{figure}

Understanding the evolution of band structure under varied temperatures is crucial for realizing the QSH effect at high temperatures.
Therefore, We further investigate the electronic properties under varied temperatures for samples on both substrates in this section, utilizing ARPES, STM, and first-principles calculations. 
Figure~\ref{figure2} presents the ARPES results of the 1T'-WSe$_2$/BLG and 1T'-WSe$_2$/SrTiO$_3$ samples.
Due to the stronger electron transfer between 1T'-WSe$_2$ and SrTiO$_3$ via interlayer effects, 1T'-WSe$_2$/SrTiO$_3$ exhibits a higher doping level compared to 1T'-WSe$_2$/BLG \cite{WSe2_interface}. This difference in doping level provides another variable in addition to the varying temperature.
\par

Since the surface lattices of BLG and SrTiO$_3$ substrates host different symmetries, the corresponding 
1T'-WSe$_2$ films with complex Brillouin zones exhibit distinct geographies.
The 1T'-WSe$_2$ grown on BLG substrate has three-fold rotated domains as shown in Fig.~\ref{figure2}(a), whereas the 1T'-WSe$_2$ grown on SrTiO$_3$ substrate has two-fold rotated domains as shown in Fig.~\ref{figure2}(b).
A series of ARPES spectra were taken from the compound Brillouin zones under varied temperatures (additional results appended in the Supplementary Information Fig. S1, S2), with LT (7 K) and RT (300 K) results presented in Fig.~\ref{figure2}(c, e, g, i) and (d, f, h, j), respectively. 
By comparing spectra between LT and RT, we found that the band structures of 1T'-WSe$_2$/BLG sample shows evident differences, while the changes in 1T'-WSe$_2$/SrTiO$_3$ sample are negligible. 
\par

\onecolumngrid
\twocolumngrid
\par

To characterize temperature effects on band structure quantitatively, energy distribution curves (EDCs) at $\Gamma$ and the momentum of the conduction band minimum (CBM) were extracted from ARPES spectra as shown in Fig.~\ref{figure3}(a, d).
More EDCs at different momentum positions are presented in the Supplementary Information Fig. S3.
For the 1T'-WSe$_2$/BLG sample, it is evident that $E_{V2}$ positively relates to increasing temperature. However, the evolution of $E_{CBM}$ is challenging to discern due to the impact of the CG and the possibility of $E_{CBM} > E_F$ at high temperatures.
Determining the fundamental gap evolutions under varied temperatures is crucial for the QSH effect. 
Meanwhile, the subbands positions also play an important role in transport properties in WSe$_2$ \cite{WSe2_subband} and MoS$_2$ \cite{MoS2_minigap}, where the twisted-resonant tunneling conductivity and negative-differential resistance are affected by the minigaps below Fermi energy. 
Therefore, we defined the fundamental gap $\Delta_1 = E_{CBM} - E_{V1}$ and the gap $\Delta_2 = E_{V1} - E_{V2}$, and plot their changes with varied temperatures in Fig.~\ref{figure3}(b). 
The $\Delta_2$ is decreasing with increasing temperature. 
In contrast, the fundamental gap $\Delta_1$ is slightly increasing with increasing temperature. 
However, the values of $\Delta_1$ above 200 K is lacking, since the $E_{CBM}$ is challenging to discern due to the impact of the CG and the possibility of $E_{CBM} > E_f$ at temperatures above 200 K.
With the current data, we conclude that the trend of $\Delta_1$ and $\Delta_2$ is positively and negatively related to increasing temperatures for the case of 1T'-WSe$_2$/BLG.
\par

\onecolumngrid
\begin{figure}[h]
\begin{minipage}[c]{\textwidth}
    \centering
    \includegraphics{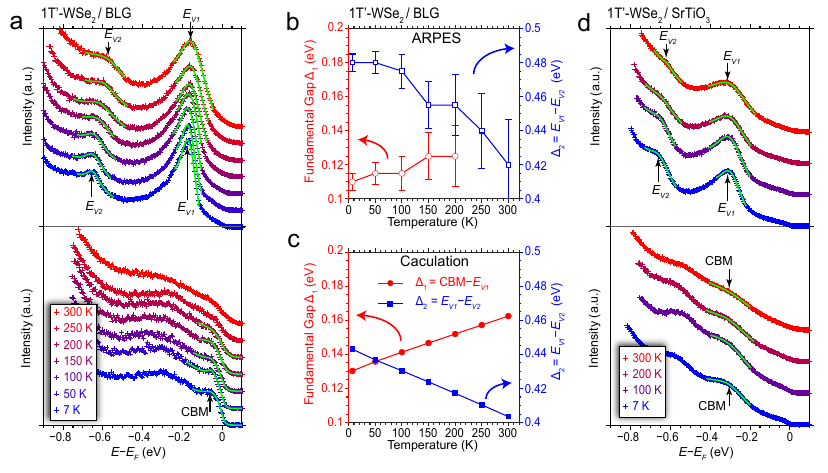}
    \caption{\label{figure3}
        \textbf{(a) and (d)}EDCs at varied temperatures at the $\Gamma$ point and CBM for (a) 1T'-WSe$_2$/BLG and (d) 1T'-WSe$_2$/SrTiO$_3$, respectively.
        The crosses are the raw data, and the green curves are the Lorentzian peak fitting on the raw data.
        The detailed fitting method can be seem in the Supplementary Information part B.3.
        \textbf{(b)}The experimental values of $\Delta_1 = E_{CBM} - E_{V1}$ and $\Delta_2 = E_{V1} - E_{V2}$ extracted from EDCs in (a). The errors are estimated as $\sqrt{(5)^2 + (k_{B}T)^2}$ meV, where the 5 meV is the resolution ability of the analyzer. 
        \textbf{(c)}The calculated values of the $\Delta_1 = E_{CBM} - E_{V1}$ and $\Delta_2 = E_{V1} - E_{V2}$ based on the lattice constants of 1T'-WSe$_2$/BLG corresponding to temperature changing.   
    }
\end{minipage}
\end{figure}
\twocolumngrid
\par

\onecolumngrid
\begin{figure}[h]
\begin{minipage}[c]{\textwidth}
    \includegraphics{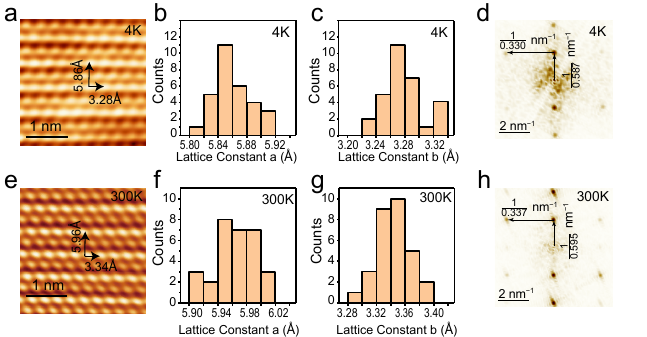}
    \caption{\label{figure4} 
        \textbf{(a) and (d)} Atomic images of 1T'-WSe$_2$/BLG with measured lattice constants at 4 K and 300 K, respectively. 
        \textbf{(b), (c), (e) and (f)} Statistical histograms of the lattice constants from 30 different positions of 1T'-WSe$_2$/BLG at 4 K and 300 K, respectively. The average lattice constants are $\bm{a}=5.86 \pm 0.03$ \AA, $\bm{b}=3.28 \pm 0.03$ \AA at 4 K, and $\bm{a}=5.96 \pm 0.03$ \AA, $\bm{b}=3.34 \pm 0.02$ \AA\ at 300 K.
        \textbf{(d) and (h)} FFT images of (a) and (e), 
    }
\end{minipage}
\end{figure}
\twocolumngrid
\par

In order to understand the gap evolution under varied temperatures, we propose that the thermal expansion affect the lattice constant of the 1T'-WSe$_2$/BLG. To verify this point, we extracted the lattice constants from both LT and RT STM data of 1T'-WSe$_2$/BLG in Fig.~\ref{figure4}.
The STM were carefully calibrated by using BLG substrate [see the Supplementary Information Fig. S4]. 
The statistical histograms of values of lattice constants ($\bm{a},\bm{b}$) taken from 30 different positions are shown in Fig.~\ref{figure4}(b,c) and (f,g).
At 4 K, the average lattice constants are $\bm{a} = 5.86 \pm 0.03$ \AA, $\bm{b} = 3.28 \pm 0.03$ \AA\, and the median values of lattice constants are $\bm{a} = 5.85 \pm 0.03$ \AA, $\bm{b} = 3.27 \pm 0.03$ \AA.
At 300 K, the average lattice constants are $\bm{a} = 5.96 \pm 0.03$ \AA, $\bm{b} = 3.34 \pm 0.03$ \AA\, and the median values of lattice constants are $\bm{a} = 5.97 \pm 0.03$ \AA, $\bm{b} = 3.35 \pm 0.02$ \AA.
The lattice constants can also extract from the fast-Fourier-transform (FFT) images shown in Fig.~\ref{figure4}(d,h).
At 4 K, the lattice constants are $\bm{a} = 5.87 \pm 0.05$ \AA, $\bm{b} = 3.30 \pm 0.05$ \AA.
At 300 K, the lattice constants are $\bm{a} = 5.95 \pm 0.05$ \AA, $\bm{b} = 3.37 \pm 0.05$ \AA.
The results from different techniques show good agreement, indicating a sizable lattice expansion at RT. 
The expansion rates from LT to RT are both larger than $1.7 \%$ and $3.0 \%$ for $\bm{a}$ and $\bm{b}$, clearly evidencing thermal expansion effect of  1T'-WSe$_2$/BLG. 
The thermal-expansion coefficient (TEC) of 1T'-WSe$_2$/BLG is approximately 60$\times$10$^{-6}$ K$^{-1}$, a value considerably large compared to other two-dimensional materials such as 1H-MoS$_2$ (approximately 14.5$\times$10$^{-6}$ K$^{-1}$ \cite{Thermal_MoS2}). 
Based on the average lattice constants at 4 K and 300 K, we carried out first-principles calculations, and the calculated band structures present good matches with the corresponding ARPES spectra as shown in Fig.~\ref{figure2}(c,d,e,f). 
Furthermore, assuming linear dependence between lattice constants and temperature, we calculated a series of band structures (appended in Fig. S5) and present the gap evolutions of $\Delta_1$ and $\Delta_2$ in Fig.~\ref{figure3}(c). 
The calculated $\Delta_1$ is positively related to increasing temperatures, $\Delta_2$ is negatively related to increasing temperatures, consistent with the trend from ARPES results, substantiating the impact of the thermal expansion effect. 
\par

However, for the case of 1T'-WSe$_2$/SrTiO$_3$, all the $E_{V1}$, $E_{V2}$ and $E_{CBM}$ show few shifts under varied temperatures in Fig.~\ref{figure2}(g,h,i,j) and Fig.~\ref{figure3}(d).
The negligible band evolution under varied temperatures in 1T'-WSe$_2$/SrTiO$_3$ can be attributed to the enhanced interface effect, as mentioned in a previous study\cite{WSe2_interface}. 
We proposed that this interface effect can effectively suppress the thermal expansions of the grown 1T'-WSe$_2$ due to the relatively small TEC of SrTiO$_3$ substrate (approximately 9.4$\times$10$^{-6}$ K$^{-1}$) \cite{SrTa_phasetransition}. 
As a result, the band structure of 1T'-WSe$_2$/SrTiO$_3$ does not show alternation around Fermi level.
\par

\subsection{Robust CG in the Epitaxial 1T'-WSe$_2$ Monolayer}

\begin{figure}[h]
        \begin{minipage}[c]{\textwidth}
        \centering
            \includegraphics{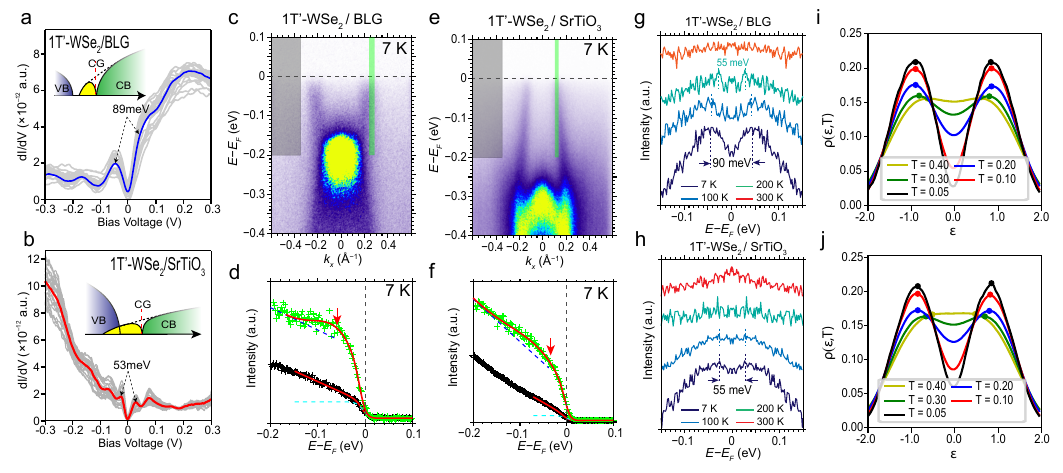}
            \caption{\label{figure5} 
                \textbf{(a) and (b)} STS dI/dV spectra of (a) 1T'-WSe$_2$/BLG (b) and 1T'-WSe$_2$/SrTiO$_3$, respectively.
                The grey lines represent dI/dV spectra taken from 20 different locations on the surface of 1T'-WSe$_2$.
                The blue and red lines denote the statistical mean values of these grey lines. 
                Insets in (a) and (b) display schematics of the band diagram around the Fermi level ($E_f$), as inferred from the corresponding ARPES and DOS data. 
                The energy position of CG is indicated by the red dashed line in these schematics. 
                All the STS spectra were taken at 4 K. 
                \textbf{(c) and (e)} ARPES spectra along the Y-$\Gamma$-Y direction of 1T'-WSe$_2$/BLG and 1T'-WSe$_2$/SrTiO$_3$, respectively, taken at 7 K.
                \textbf{(d) and (f)} EDCs at the background regions [gray areas in (c) and (e)], plotted as back crosses, and at the band intersecting Fermi level [green areas in (c) and (e)], plotted as green crosses, respectively.
                The red curves are the fitting results of a Fermi function with linear background (with a Lorentzian peak for the band EDC).
                \textbf{(g) and (h)} Symmetrized EDCs with the Fermi function background removed at temperatures of 7 K, 100 K, 200 K, 300 K. 
                \textbf{(i) and (j)} Numerical simulation results of temperature-dependent DOS of a 2D 16$\times$16 lattice system by Monte Carlo, with different screening radii: $r_0 = 10^{5}A$ and $r_0 = 10^{1}A$, respectively, and the temperature in unit $\frac{e^2}{A\kappa k_B}$, where $A$ is the lattice constant and $\kappa$ is the dielectric constant.
                }
        \end{minipage}
    \end{figure}

In the varied-temperature investigation of 1T'-WSe$_2$ electronic properties, we combined in-situ STS and ARPES to investigate the band structure and DOS near Fermi level. 
In the STS shown in Fig.~\ref{figure5}(a,b), the V-shaped dips pinned at Fermi level was observed in both cases of 1T'-WSe$_2$/BLG and 1T'-WSe$_2$/SrTiO$_3$.
We inferred these dips to be the CG as previously reported in 1T'-WTe$_2$\cite{WTe2_coulomb,WTe2_strain}.
To ensure the discovery of the CG in our experiments, STS data were taken from 20 different locations (along the black dashed lines in Fig.\ref{figure1}(c, d)) and all exhibit the V-shaped dip at Fermi level depicted in gray lines in Fig.~\ref{figure5}(a, b). 
This noticeable reduction in DOS is not a result of the averaging process but a universally observed phenomenon in 1T'-WSe$_2$, which is also reflected in the spatial STS spectra in the Supplementary Information Fig. S6.
To evaluate the value of CGs of 1T'-WSe$_2$/BLG and 1T'-WSe$_2$/SrTiO$_3$, we averaged the STS data and present them in blue and red lines in Fig.~\ref{figure5}(a) and (b).
Approximately 89 meV and 53 meV CGs were observed in 1T'-WSe$_2$/BLG and 1T'-WSe$_2$/SrTiO$_3$, respectively (the statistics of CGs is shown in the Supplementary Information Fig. S7).
\par

The existence of CG was also confirmed in the ARPES spectra near Fermi level as shown in Fig.~\ref{figure5}(c,e). The Fermi level was carefully calibrated by the graphene/SiC and potassium film samples (see the Supplementary Information Fig. S8). In Fig.~\ref{figure5}(d,f), the EDCs at the background regions [gray areas in Fig.~\ref{figure5}(c,e)] are plotted as the black crosses, and the EDCs at the momenta that conduction bands intersect Fermi level are plotted as the green crosses. For the background EDCs, we applied the Fermi function with linear background fitting to the raw data, and the results are plotted as the red curves (the Fermi function is plotted as the green dashed curves). In contrast, the fitting results to the EDCs of the bands show significant spectral weight transfer at Fermi level (marked by the red arrows), indicating the CGs in both 1T'-WSe$_2$/BLG and 1T'-WSe$_2$/SrTiO$_3$.
\par

The observed CG size of 1T'-WSe$_2$/BLG is distinctly larger than that of in 1T'-WSe$_2$/SrTiO$_3$. 
Based on the integrated ARPES spectra with first-principles calculations shown in Fig.~\ref{figure2},the strong interface effect in 1T'-WSe$_2$/SrTiO$_3$ provides large charge transfer. 
The gap size in 2D system is dependent on the dielectric constant ($\kappa$): $\propto g_0^{0.5}/\kappa^{1.5}$\cite{CoulombGap2}, which is decreasing with the heavier doping\cite{CoulombGap3}.
In this reason, the 1T'-WSe$_2$/SrTiO$_3$ show a smaller CG size than the 1T'-WSe$_2$/BLG.
\par

\begin{figure}[h!]
        \begin{minipage}[c]{\textwidth}
        \centering
            \includegraphics{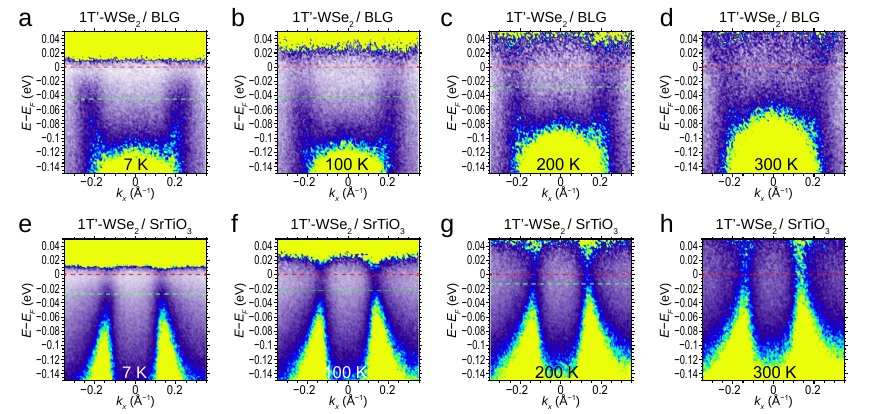}
            \caption{\label{figure6} 
                ARPES spectra divided by the corresponding Fermi-Dirac distribution functions.
                \textbf{(a-d)} The spectra of 1T’-WSe$_2$/BLG at (a) 7K, (b) 100 K, (c) 200 K and (d) 300 K, respectively.
                \textbf{(e-h)} The spectra of 1T’-WSe$_2$/SrTiO$_3$ at (e) 7K, (f) 100 K, (g) 200 K and (h) 300 K, respectively.
                The red dashed lines are Fermi level. The green dashed lines are the gap peak position extracted from the fitting of EDCs.
                }
        \end{minipage}
    \end{figure}

We further investigated the evolution of the CGs under varied temperatures by ARPES.
To evaluate the CGs sizes from the ARPES results, we divide the EDCs of the band [green crosses in Fig.~\ref{figure5}(d,f)] by the Fermi background [black crosses in Fig.~\ref{figure5}(d,f)], and symmetrize them at Fermi level. The symmetrized EDCs at varied temperatures are shown in Fig.~\ref{figure5}(g,h).
The CGs at Fermi level whose sizes estimated from the symmetrized EDCs at 7 K agree with the STS results at 4 K.
In both 1T'-WSe$_2$/BLG and 1T'-WSe$_2$/SrTiO$_3$, the width and depth of CGs gradually decrease with increasing temperatures.
For the 1T'-WSe$_2$/BLG sample, the CG can persist up to 200 K with estimated sizes of 55 meV.
For the 1T'-WSe$_2$/SrTiO$_3$ sample, the CG can be resolved up to 100 K from the the symmetrized EDCs.
We also plot the ARPES spectra divided by the corresponding Fermi-Dirac distribution functions at varied temperatures in Fig.~\ref{figure6}, and the CGs can also be well directly resolved in these spectra.
We found that the suppression of DOS at Fermi level due to the CG can persist up 200 K in both 1T'-WSe$_2$/BLG and 1T'-WSe$_2$/SrTiO$_3$, indicating the robustness of the CG.
The temperature dependence of the CG sizes coincide with the results of previous theoretical study \cite{CoulombGap_temperature}.
\par

\onecolumngrid
\twocolumngrid

To further study the effects of varied temperature and doping level on CG, we performed Monte Carlo simulations considering different screenings and temperatures. 
The simulation model primarily accounts for the Coulomb interaction, contributing to the decreasing DOS at Fermi level \cite{CoulombGap1,CoulombGap2} (see Supplementary Information part B.3 for a detailed description).
We accounted for screening and temperature effects by different screening radii in Coulomb potential: $e^{-\frac{r_{ij}}{r_0}}r_{ij}^{-1}$ and the update acceptance condition: $e^{\frac{\delta E}{k_B T}}$ in simulation. 
Based on the acceptance condition, critical constant can be defined as $T_{\kappa}/T, T_{\kappa} = \frac{e^2}{A\kappa k_B}$ to analyze the CG behavior under varied temperatures.
The simulation results show disappearing dip trend with increasing temperatures for two different doping levels, as presented in Fig.~\ref{figure5}(i, j). 
The effects of temperature and doping are superposed, both suppressing the Coulomb effect in 1T'-WSe$_2$ (see the trend of doping effect in Fig. S6).
The result from heavier doping levels coincide with the weaker and smaller CG in 1T'-WSe$_2$/SrTiO$_3$.
Combined the robustness of CG in 1T'-WSe$_2$, the value of $T_{\kappa}/T$ is supposed to be small, which can come from the relative large $\kappa$\cite{Dielectric_wte2}. The condition of $T_{\kappa}/T <<1$ coincides the low-temperature CG behavior in our simulation.

In conclusion, our numerical simulation provides evidence that the weaker and smaller CG in 1T'-WSe$_2$/SrTiO$_3$ can be attributed to the heavier electron doping, and both intensified screening and increased temperature mediate the CGs.
Combined with experimental data at varied temperatures, the CG survives can survive up to 200 K in  1T'-WSe$_2$/BLG due to the less electron doping, and we conclude that the Coulomb effect is considerably robust, aiding the insulating bulk behavior in 1T'-WSe$_2$.
\par
\newpage

\onecolumngrid
\twocolumngrid

\section{Discussions}

In this letter, the temperature effect is systematically studied in epitaxial 1T'-WSe$_2$/BLG and 1T'-WSe$_2$/SrTiO$_3$. 
A significant thermal expansion effect is found in 1T'-WSe$_2$/BLG, while the lattice of 1T'-WSe$_2$/SrTiO$_3$ shows no significant lattice changes. 
The TEC of 1T'-WSe$_2$/BLG is estimated to be approximately 60$\times$10$^{-6}$ K$^{-1}$, much larger than that of the 1H phase\cite{Thermal_MoS2}. 
The expansion of the lattice is beneficial to the insulating bulk properties, where our calculation predicts larger fundamental gaps at increasing temperatures.
For the case of 1T'-WSe$_2$/SrTiO$_3$, we conclude that the enhanced interlayer effect \cite{WSe2_interface} and the relatively small TEC of SrTiO$_3$ (approximately 9.4$\times$10$^{-6}$ K$^{-1}$) \cite{SrTa_phasetransition} contribute to the negligible thermal expansion in 1T'-WSe$_2$/SrTiO$_3$, and as a result, the band structure does not show alternation around Fermi level.
Therefore, We conclude that the thermal expansion effect is a significant component in varied-temperature studies, and our temperature-dependent study is essential to the application of 1T'-WSe$_2$ films in transport.
\par

Our another important finding is the Robust CGs in both 1T'-WSe$_2$/BLG and 1T'-WSe$_2$/SrTiO$_3$. The CG sizes exhibit temperature dependencies in both EDCs data and numerical simulations, with CGs decreasing gradually with increasing temperatures. 
For the 1T'-WSe$_2$/BLG, the CGs continue to persist up to 200 K, indicating the robustness of the CG.
We also observed a relatively smaller size of CG in 1T'-WSe$_2$/SrTiO$_3$ in both ARPES and STS spectra. From numerical simulation and experimental evidence, we conclude that heavier electron doping, another effect of the enhanced interlayer effect, induces a larger screening effect in 1T'-WSe$_2$/SrTiO$_3$. 
Our findings are consistent with results reported for another 1T' structure \cite{WTe2_coulomb}. Furthermore, the robustness of CG suggests a relatively strong Coulomb effect in 1T'-WSe$_2$. 
Our results provide solid evidence from both experiments and numerical simulation that 1T'-WSe$_2$ is a promising candidate for achieving QSH and open up avenues for the development of spintronics operating at higher temperatures.
\par

\section{Methods}
 
The growth of 1T'-WSe$_2$ films was carried out in an ultra-high vacuum (UHV) system equipped with a combined MBE-STM-ARPES setup at Nanjing University.
The base pressure of the system was approximately 1.5$\times$10$^{-10}$ mbar. 
Alternatively, in-situ growth of 1T'-WSe$_2$ films was also performed using an UHV MBE-STM system (Unisoku Co. USM1300) at Nano-X, Suzhou Institute of Nano-Tech and Nano-Bionics (SINANO).
The bilayer graphene (BLG) substrates were prepared by flash-annealing 4H-SiC (0001) wafers at 1300 $^{\circ}$C for 60$\sim$80 cycles \cite{BLG}. 
The SrTiO$_3$(100) substrate was prepared by direct heating at 950$^{\circ}$C for 1 hour.
For the deposition of WSe$_2$, a high purity (99.95\%) tungsten rod was used as the source material and evaporated using an electron-beam heating evaporator with a flux monitor function. 
The selenium (Se) precursor, with a purity of 99.9995\%, was evaporated from a standard Knudsen cell.
During the growth process, the BLG substrate was kept at 250 $^{\circ}$C, and the SrTiO$_3$(100) substrate was kept at 450 $^{\circ}$C to promote the formation of the 1T' phase of WSe$_2$ monolayer \cite{WSe2_transition5,WSe2_interface}.
\par

The in-situ ARPES measurements were conducted using a SCIENTA DA30L analyzer. A monochromatic Helium lamp operating in He I mode (21.2 eV) was employed as the excitation light source.
To measure the spectra at different temperatures in ARPES, a Helium-free close-cycle cryo-manipulator was utilized, allowing the temperature to be varied from 7 K to 300 K. 
The energy resolution of the DA30L analyzer was approximately 5 meV, and the total energy resolution in the ARPES measurements was approximately $\sqrt{(5)^2 + (k_{B}T)^2}$ meV.
\par

The room-temperature (RT, 300 K) STM measurements were conducted in a in-situ Pan-style STM at Nanjing University.
The low-temperature (LT, 4 K) STM/STS measurements were conducted at Nano-X, Suzhou Institute of Nano-Tech and Nano-Bionics (SINANO), China, using the Unisoku Co. USM1300. 
STS differential conductance (dI/dV) point spectra and spatial mappings were acquired in constant-height mode employing standard lock-in techniques with a frequency of 973 Hz and an Vrms of 10 mV. 
The dI/dV spectra on Ag (111) served as a reference for STS measurements. 
The STM/STS data were analyzed using WSxM software.
\par

The band structure calculations were conducted within the density functional theory framework using the VASP package \cite{VASP_2}. 
The projector augmented wave method was employed \cite{pseudopotentials_PAW}.
The Perdew-Burke-Ernzerhof exchange-correlation functional within the generalized-gradient approximation was used \cite{PBE}. 
To improve the accuracy of band gap calculations, we adapted the HSE03 hybrid functional \cite{DFT_HSE}. The band structure along high-symmetry lines was obtained from the WANNIER90 interface tight-binding model \cite{Maximal_localized_Wannier}.
In the varied-temperatures simulation, only thermal expansion effect is considered and the electron-phonon coupling is not included in calculation.
For the simulation of the CG, a two-dimensional squared lattice model incorporating the Coulomb interaction was employed \cite{CoulombGap1,CoulombGap2}. 
The density of states at different energies was evaluated, taking into account temperature and screening effects. Further details regarding the calculation can be found in the supplementary materials.

\begin{acknowledgments}
This work is supported by the National Natural Science Foundation of China (Nos. 92165205, 12134008, 12204512), the Innovation Program for Quantum Science and Technology of China (Nos. 2021ZD0302803), the National Key Research and Development Program of China (No. 2018YFA0306800), the Suzhou Science and Technology Program (No. SJC2021009), the Hong Kong Research Grants Council (ECS26302118, 16305019, and N\_HKUST626/18), and the fundamental Research Funds for the Central Universities (No. 2024/14380228).
\end{acknowledgments}

\bibliography{main}

\end{document}